# A Survey on Metaverse: the State-of-the-art, Technologies, Applications, and Challenges

Huansheng Ning，Hang Wang, Yujia Lin, Wenxi Wang, Sahraoui Dhelim,
Fadi Farha, Jianguo Ding, Mahmoud Daneshmand

Abstract:
Metaverse is a new type of Internet application and social form that integrates a variety of new technologies. It has the characteristics of multi-technology, sociality, and hyper spatiotemporality. This paper introduces the development status of Metaverse, from the five perspectives of network infrastructure, management technology, basic common technology, virtual reality object connection, and virtual reality convergence, it introduces the technical framework of Metaverse. This paper also introduces the nature of Metaverse's social and hyper spatiotemporality, and discusses the first application areas of Metaverse and some of the problems and challenges it may face.

# 1 Introduction

Metaverse is a new type of Internet application and social form that integrates a variety of new technologies. It provides an immersive experience based on augmented reality technology, creates a mirror image of the real world based on digital twin technology, builds an economic system based on blockchain technology, and tightly integrates the virtual world and the real world into the economic system, the social system, and the identity system, allowing each user to produce content and edit the world. Metaverse is still a concept that is constantly evolving, and different participants are enriching its meaning in their own ways.

From a technical perspective, the way humans communicate is constantly improving (see Figure1). Accordingly, technological innovation, the integration of several new technologies, and new Internet applications have also developed.

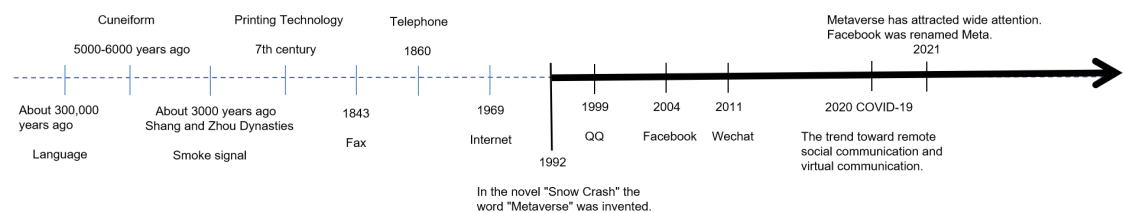

Figure 1 Timeline of the development of communication methods

From a commercial perspective, Metaverse has shown far-reaching commercial prospects. As a new capital export, large companies have devoted themselves to building the Metaverse. From the user's point of view, the demands on the freedom of the virtual world, the content and interaction methods of the Internet, etc. are constantly increasing.

This paper summarizes the development status of the Metaverse, starting from the definition of the Metaverse, and presents the technical framework, the social framework, and the hyperspace. It also predicts and discusses the future development of the Metaverse.

Section 2 introduces the development status of Metaverse. Section 3 introduces the three characteristics of Metaverse: multi-technology dominance, sociality, and hyperspace. Section 4, based on the multi-technology dominance of the Metaverse, introduces a variety of new technologies and the framework of the Metaverse. Section 5 based on the social nature of Metaverse, introduces Metaverse as a new social form, including economic system, cultural civilization system, legal system and other aspects. Section 6 based on the time-space nature of Metaverse, further introduces the free, immersive, and multi-dimensional nature of Metaverse. Section 7, based on the current development and planning of Metaverse, predicts the first application field of Metaverse. Section 8 discusses several problems that need to be solved in Metaverse. Section 9, the conclusion part, summarizes the development of this paper and Metaverse.

## 2 Recent advances of the Metaverse

This section focuses on the policies of different countries and international organizations, representative enterprises and their typical products, combined with the statistics on the number of Metaverse-related publications in the Web of Science database and SCOPUS database, and analyzes the development history and current status of the Metaverse. Analyze the relationship between each country's Metaverse policy and the design of the country's Metaverse enterprises.

## 2.1 The policy of national and international organizations

Different countries and international organizations have different attitudes towards the Metaverse. In this section, we will analyze the development status of Metaverse by examining the policies of different countries and international organizations from the aspects of politics, economics, and government attitudes.

Because in the Metaverse economic system, the most important foundation is the token economy based on the Blockchain. Blockchain is the most important technology of digital currency. Therefore, we will analyze the policies of different countries and international organizations regarding the Blockchain and digital currency in terms of economy. Artificial intelligence, interactive technology, cloud computing and edge computing are the supporting technologies of the Metaverse. Therefore, we will analyze the policies of different countries and international organizations regarding the development of these technologies in terms of technology.

The specific national policies are listed in Table 2.1.

**Table 2.1 The specific national policies**

| | | |
|---|---|---|
| USA | Government | ONC, Office of the National Coordinator for Health Information Technology, a division of the U.S. Department of Health and Human Services, organized a marathon of application development by healthcare hackers to apply blockchain technology to the healthcare sector. The executive branch of Congress, on behalf of the **Trump** administration, has recognized the potential of blockchain and called for the development of blockchain technology in the public sector. |
| | Economy | The US Congress announced the establishment of the Congressional Blockchain Decision Committee.<br>The "Statement on Potentially Illegal Crypto Asset Trading Platforms" issued by the United States Securities and Exchange Commission (SEC) confirms that digital encrypted assets belong to the scope of securities, but do not have the specific attributes and legal status of an actual currency. |
| China | Economy | In December 2016, the state included blockchain technology in the "13th Five-Year" national informatization plan.<br>In October 2016, the Ministry of Industry and Information Technology released the "2016 China Blockchain Technology and Application Development White Paper".<br>In May 2021, the Ministry of Industry and Information Technology issued the "Guiding Opinions on Accelerating the Application and Industrial Development of Blockchain Technology". |
| | Technology | A spokesman for the Ministry of Industry and Information |
| | | Technology said that the deployment of 5G will accelerate in 2020, with more than 600 thousand new 5G base stations opening during the year. The penetration rate of 5G is in a phase of rapid improvement. It is estimated that by 2023, the penetration rate of 5G individual users will exceed 40%, the number of users will exceed 560 million, and the proportion of 5G network access traffic will exceed 50%.<br>The State Council has issued the policy document "Measures for Evaluating the Security of Cloud Computing Services". |
| Japan | Economy | The Japanese government has established the first blockchain industry organization - the Japan Blockchain Association (JBA) - and a blockchain cooperative alliance.<br>On April 1, 2017, Japan enacted the "Payment Services Act", which officially recognizes Bitcoin as a legal payment method and sets clear regulatory requirements for the exchange of digital, encrypted assets.<br>The Japanese economic authorities have defined the "Metaverse", but they do not considered it as a definitive business form for now. The ministry plans to improve laws and development guidelines and is trying to take a leading position in the global virtual space industry. |
| South Korea | Government | The Korean government hopes to play a leading role in the Metaverse industry.<br>The government plans to allocate 9.3 trillion won (about 51.6 billion yuan) by 2022 to accelerate digital transformation and cultivate new industries such as Metaverse, as well as help small and medium-sized enterprises and venture capital firms build blockchains.<br>In 2025, 2.6 trillion won (about $2.2 billion) is expected to be spent on Metaverse, blockchain, and other technologies.<br>From 2022 to 2025, the Metaverse platform developed by Korean companies will be fully supported. |
| | Technology | The Korean Office of Technology and Standards is striving to become the world's technology leader in the Metaverse.<br>Li Xianghong, head of the Korean Office of Technology and Standards, said at the online meeting that the Korean government will provide the necessary support to any company that wants to formulate technical standards in this field. |
| | Education | The South Korean Ministry of Education is promoting lessons in the Metaverse.<br>The Seoul Education Bureau said it would open relevant courses in the city gather to open the door to the world of the Metaverse to 2100 primary and secondary school students. Kim Enbei, director of Seoul Education Bureau, said, "By operating the creative science classroom based on the Metaverse, we offer virtual reality and artificial intelligence, and try our best to become the center of integrated |

| | | science education." |
|---|---|---|
| United Arab Emirates | Economy | In 2016, Dubai established the Global Blockchain Committee.<br>In 2017, the Dubai government announced that Emcredit, a subsidiary of Dubai Economy, would cooperate with the US-based startup Object Tech Grp Ltd to create an encrypted digital currency called emCash.<br>In late 2020, the UAE Securities and Commodities Authority (SCA) issued the "Chairman of the Authority's 2020 Decision (21/RM) on the Regulation of Crypto Assets".<br>In October 2021, the Dubai government will host the Blockchain Summit, a mutually beneficial initiative to transition from a regulated crypto economy to a Metaverse.<br>The UAE Central Bank announced a planning roadmap for the period 2023-2026 to realize the operation of the CBDC and make the UAE one of the top ten countries in the world for the digital transformation of the financial sector. |
| | Technology | At the 12th Global Mobile Broadband Forum (MBBF), the CEO of the UAE's SAAMENA Communications Commission stated that all Gulf countries have released commercial 5G networks. |

The specific guidelines of international organizations are listed in Table 2.2

**Table 2.2 The specific guidelines of international organizations**

| | |
|---|---|
| ITU | In August 2020, the ITU published the first set of international standards for blockchain.<br>From April 19 to 30, 2021, the ITU will hold a plenary session. At the meeting, three proposals for international blockchain standards led by the Chinese Academy of Information and Communications Technology were approved.<br>From February 23 to March 2, 2016, the 23rd meeting of the ITU Working Group was held in Beijing. The Chinese Academy of Information and Communications Technology was the Chinese sponsor. The ITU is beginning to evaluate 5G technology and explore new spectrum for mobile communications. |
| IEEE | On December 6, 2017, the establishment of the IEEE Blockchain Asset Trading Committee was formally approved.<br>On December 23, 2020, IEEE 2418.2-2020 "IEEE Standard for Data Format for Blockchain Systems" was officially published and implemented under the leadership of China Electronics Standardization Institute<br>IEEE released the "IEEE Global CIO and CTO Interview Survey: Opportunities and Challenges in 2021, and Key Technology Trends." According to the survey results, artificial intelligence and machine learning, 5G and Internet of Things technologies will become the most important technologies influencing 2021. |
| IET | IET Blockchain was launched in collaboration with the IET, Tongji University and Shanghai Blockchain Application SERCBAAS. It aims to publish cutting-edge results related to blockchain basic theories, applied technology and industrial development, technological innovation and the latest viewpoints. |
| WWW | On June 8-9,, 2016, the first Decentralized Web Summit was held in San Francisco, USA. Tim Berners-Lee, the inventor of the WWW, mentioned at the conference that blockchain and P2P technology will be used to create a decentralized Internet. World Wide Web Consortium W3C held a blockchain seminar, calling for the creation of public standards for blockchain technology. |

## 2.2 Representative companies in different countries and their typical products/services

Metaverse integrates the most advanced technologies such as 5G, cloud computing, computer vision, blockchain, artificial intelligence, etc., and has applications in numerous fields such as video games, art, and business. Through the research in Section 2.1, we have a preliminary understanding of the different policies of international organizations and countries on the Metaverse. Due to the different policies of different countries in the Metaverse, the representative companies and their typical products and development plans are also different in different countries. For example, the United States, as the pioneer of Metaverse, has a relatively extensive Metaverse layout, which is applied in many fields such as business, games, arts and social affairs. China has a large market and strong Internet enterprises and Internet applications. Domestic Internet companies have successively introduced business, video games, and art in the Metaverse. Japan, with its cumulative advantages in the ACG industry and rich IP resources, is focusing on its application areas in animation and video games, while South Korea is government-led and driven by the idol industry. German and Italian luxury brands are trying to make more people their customers through virtual products, etc. Other representative companies from other countries and their typical products are listed in Table 2.3.

Table 2.3 Representative companies in different countries and their typical products/services

| | | |
|---|---|---|
| The USA | Amazon | Since 2018, Amazon has been developing a "new VR shopping experience" and trying to use Metaverse to create a virtual shopping space where shoppers can interact with digital products by building a kind of virtual "Amazon shopping mall" in Metaverse to assert its dominant position in the market. |
| | Roblox | Players can create their own virtual world or write various games, imagination is the only limit. Roblox supports VR devices to enhance user immersion. Roblox has become the world's largest game UGC platform, supporting iOS, Android, PC, Mac |
| | | and other platforms. Roblox is currently one of the "worlds" that have the potential to be closest to the Metaverse. |
| | Facebook | In September 2019, Facebook released the VR social platform Facebook Horizon, and launched a public beta in August 2020.<br>In July 2021, Facebook announced that it will become the Metaverse team, transforming into a Metaverse company within five years and investing at least $10 billion in its Reality Labs project.<br>In October 2021, Facebook announced that it was renaming itself Meta. |
| | Epic Games | n April 2021, Epic Games announced a $1 billion investment to build a metaverse. And acquired Skethfab, the largest platform for 3D models, to absorb user traffic from the Skethfab platform and increase its market share in the metaverse. |
| | Disney | Tilak Mandadi, Disney's chief technology officer, said building a "theme park metaverse" will be the next step in the evolution of Disney's theme parks. |
| | Snapchat | Snapchat has introduced custom avatars and filters to fill the world with digital content.<br>Currently Snapchat has launched the Bitmoji service, which allows users to pose in physical snapshots and create their own 3D Bitmoji avatars. |
| | Nvidia | On August 11, 2021, Nvidia announced the Nvidia Omniverse plan to create the world's first virtual collaboration and simulation platform. |
| | Microsoft | Microsoft is cautious about Metaverse. Wei Qing: The "Metaverse" has practical value only when it returns to the physical world. |
| | Decentraland | VR virtual world based on Ethereum, the first fully decentralized virtual world owned by users. The core content of Decentraland is artwork, and there is a place dedicated to the exhibition of digital artwork. |
| China | Tencent | Tencent has made a whole series of investments in the Metaverse ecosystem, including the AR development platform, the "Avakin life" game, the Spotify music streaming platform, etc., and applied for registration of the "Kings Metaverse" and "TiMi Metaverse" trademarks in September 2021. |
| | Alibaba | Alibaba applied for the registration of trademarks such as "Ali Metaverse" and "Taobao Metaverse".<br>Tan Ping, the person in charge of XR Lab, divides the Metaverse into four layers: L1 (holographic construction), L2 (holographic simulation), L3 (virtual and real fusion), L4 (virtual and real linkage). |
| | ByteDance | ByteDance owns high-traffic platforms such as Douyin, has also |
| | | invested in visual computing and AI computing platform Moore Thread, released the game "Restart the World", and acquired PICO, a Chinese VR equipment company. |
| | NetEase | NetEase's layout of Metaverse focuses on the game business and provides low-threshold tools for game development.<br>The company invested in IMPROBABLE's meta-computing platform to enable third parties to build virtual worlds, and in the IMVU virtual character platform. |
| | Shenzhen Zqgame Co.,Ltd | Zqgame is a trend-setting Chinese game studio. On September 6, 2021, Zqgame released the preview of the game Brew Master. This game allows players to start businesses in a simulated environment and experience the real-life impact. |
| | Wondershare Technology Group Co.,Ltd | Wondershare Technology has invested in Realibox to enhance its business layout in the AR/VR field and provide a solid technical foundation for the initial deployment of Metaverse. |
| Japan | Sony, Hassilas | Mechaverse is the first Japanese Metaverse platform. Companies can quickly conduct product launches on this platform and provide video introductions and 3D model experiments for participants. |
| | GREE | GREE operates the Metaverse business through its subsidiary REALITY. It is estimated that 10 billion yen will be invested by 2024 to develop more than 100 million users worldwide. |
| | Avex Business Development、Digital Motion | Avex Business Development and Digital Motion established the "Virtual Avex Group", which plans to promote existing animation or game characters, host virtual artist activities, and virtualize concerts by real artists and other activities. |
| South Korea | SAMSUNG | SAMSUNG has launched the "Samsung Global Metaverse Fund". |
| | SK Telecom | In July 2021, SK Telecom introduced a virtual world called "ifland", where users can host and participate in meetings through cartoon characters. |
| | Urbanbase | Urbanbase is a 3D spatial data platform for real estate and interior design development. The company raised 13 billion won (about 450 million yuan) in the B+ funding round. The funds will be used to develop the VR/AR and 3D technologies Urbanbase needs to enter the Metaverse. |
| | Metaverse Alliance | The Korean Information and Communications Industry Promotion Agency has formed 25 organizations and companies into the "Metaverse Alliance" to build the Metaverse ecosystem under the leadership of the private sector through government and business collaboration, realizing an open Metaverse platform in various fields of reality and virtuality. |
| England | Sotheby's | British auction house Sotheby's launched the "Sotheby's Metaverse" section and held a special auction titled "Natively |
| | | Digital 1.2: The Collectors," which featured 53 pieces from NFT art collections on display at one time. |
| | Maze Theory | The famous British VR studio Maze Theory will create a "fan Metaverse" around well-known IPs and well-known fan universes. |
| United Arab Emirates | MetaDubai | MetaDubai is building a Metaverse city in Dubai based on blockchain, NFT, AI, decentralized data storage to develop the most complete virtual world image, economic system, and applications possible. |
| | Ripple | Blockchain payment company Ripple has announced the establishment of a regional headquarters at the Dubai International Financial Center (DIFC). |
| French | Stage11 | French Metaverse music platform Stage11 has closed a €5 million funding round led by European venture capital fund Otium Capital to create immersive Metaverse music. |
| German | RIMOWA | German luxury luggage brand RIMOWA announced on Instagram in May that it would cooperate with design studio NUOVA to launch 4 NFT artworks called "Blueprints from the Metaverse." |
| Italian | Gucci | Italian luxury brand Gucci has launched virtual sports shoes. After purchasing the shoes, consumers can use them in the Gucci APP and the VR social platform VR CHAT, or try them out on the game platform Roblox. |

## 2.3 Number of the Metaverse related publications

Web of Science is a web-based multi-disciplinary literature database. It is the world's largest comprehensive academic information resource covering the most disciplines. It contains a variety of core academic journals that are the most influential in various research fields such as natural sciences, engineering technology, and biomedicine. Therefore, we choose to count the number of the Metaverse-related publications published on this database to analyze the development process of Metaverse.

As of November 1, 2021, a total of 211 publications related to Metaverse have been published in this database. From Figure 2.1, the number of publications over the years, we can see that we can divide the development of the Metaverse into four stages: Embryonic Stage, Primary Stage, Ebb Stage, and Development Stage.

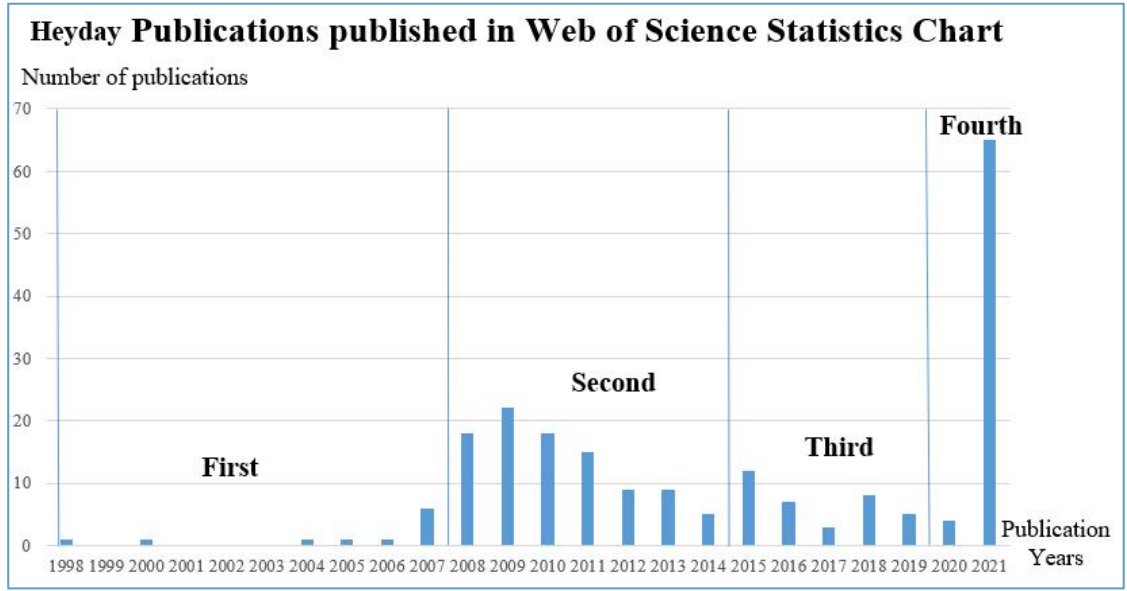

**Figure 2.1 Publications in Web of Science Statistics Chart**

The original concept of the metaverse comes from the 1992 science fiction novel "Avalanche" by Neil Stephenson, and the first publication about the metaverse on the Web of Science was published in 1998. The publication analyzed the current state of the real-time virtual human in the novel "Avalanche," including appearance, clothing and accessories, body actions, etc., opening the curtain for the study of the Metaverse. "The Matrix" from 1999 showed a real world controlled by a computer system with artificial intelligence called "Matrix". In the absence of technical tools, literature and art were used as vehicles for the study of the Metaverse in the embryonic phase, e.g. films, books, etc.

At the beginning of the 21st century, the rapid development of virtual reality technology and computer graphics created the technical basis for the development of the Metaverse. At the primary level, video games have become a new vehicle for exploring the metaverse. The game Roblox, released by Roblox in 2006, and the development of Minecraft by Mojang Studios in Sweden in 2009 greatly inspired the first wave of discussions about the Metaverse.

The development of Metaverse is still in its infancy, and its business model is not mature. Due to the open issues such as interaction issues, computing power pressures, ethical constraints, privacy risks, and addiction risks in the different worlds, and the fact that Metaverse development is still limited by current technology, research interest in Metaverse is at an Ebb Stage after 2013.

On March 10, 2021, Roblox, a sandbox game platform, included the concept of "Metaverse" in its prospectus for the first time and successfully landed on the New York Stock Exchange. The company's market value exceeded $40 billion on its first day of listing. This phenomenon sent the technology and capital worlds into an uproar and reignited the discussion about the metaverse. This year can be called the first year of Metaverse. In the development phase, Metaverse has integrated 5G, cloud computing, computer vision, blockchain, artificial intelligence and other cutting-edge science and technology that are in a phase of rapid development. Metaverse has been

applied to many fields such as medical treatment, video games, art, business, etc., and has completed its transformation from games and beyond games. The Metaverse has unprecedented explosive power.

To validate the classification of metaverse development stages by the number of metaverse-related publications published in the Web of Science database in this section, we also examined the number of metaverse-related publications published in the SCOPUS database. The results are shown in Figure 2.2:

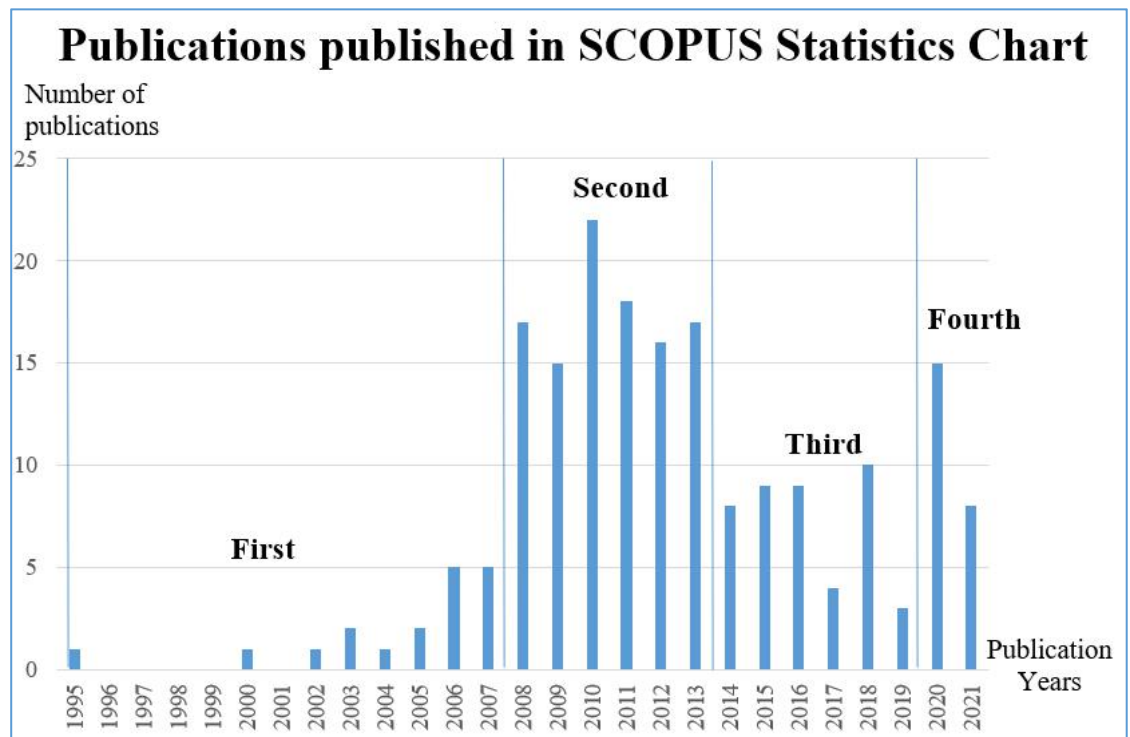

**Figure 2.2 Publications in SCOPUS Statistics Chart**

    A total of 191 publications on the Metaverse have been published in the SCOPUS database. The data obtained from the investigation in this database once again confirms our idea. According to the number of publications on the Metaverse, the development of the Metaverse can be divided into four stages, as shown in Table 2.4:

**Table 2.4 The development stage division of the Metaverse**

| Period | Stage | Description | Total number of publications in web of science | Total number of publications in SCOPUS | According to the year average, number of publications in web of science | According to the year average, number of publications in SCOPUS |
|---|---|---|---|---|---|---|
| 1992-2007 | Embryonic Stage | Mainly based on literature and art. | 11 | 18 | 0.85 | 1.39 |
| 2008- | Primary Stage | Mainly based | 91 | 105 | 15.17 | 17.5 |
| 2013 | | on video games. | | | | |
| 2014-2019 | Ebb Stage | Restricted by many open issues. | 40 | 43 | 6.67 | 7.17 |
| 2020-2021 | Development Stage | Integrate multi-technology to achieve multifield applications. | 69 | 25 | 34.5 | 12.5 |

# 3 Characteristics of Metaverse

As a new Internet application, Metaverse integrates a variety of new technologies and has the characteristics of multi-technology; as a new social form, Metaverse has the characteristics of sociality; as a parallel and closely related to the real world In the virtual world, Metaverse has the characteristics of hyper spatiotemporality.

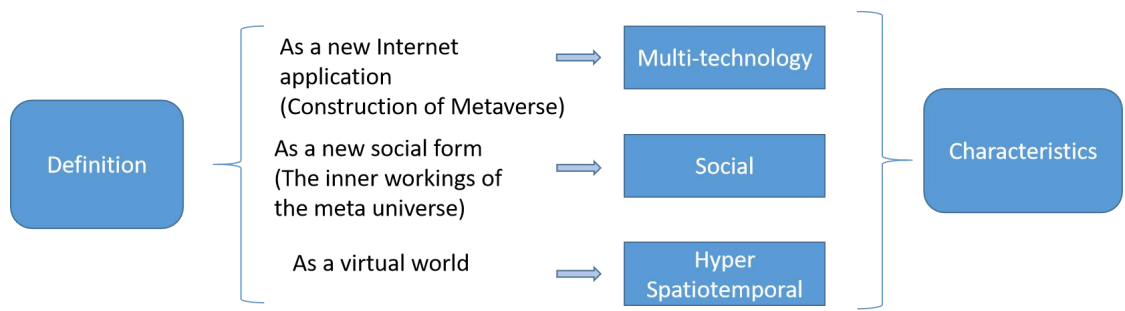

Figure 3.1 Characteristics of Metaverse

## 3.1 Multi-technology

Metaverse integrates a variety of new technologies. It provides an immersive experience based on augmented reality technology, generates a mirror image of the real world based on digital twin technology, and builds an economic system based on blockchain technology.

## 3.2 Sociality

As the definition says, the Metaverse is a new type of social form. Metaverse includes economic systems, cultural systems, and legal systems, which are closely related to reality, but have their own characteristics.

## 3.3 Hyper Spatiotemporality

Hyper Spatiotemporality refers to the Metaverse, a virtual world parallel to the real world. It breaks the boundaries of time and space and offers users an open, free and immersive experience.

# 4 Metaverse: Multi-technology convergence perspective

The Metaverse is the tight integration, interaction, and intertwining of the real and virtual worlds that requires the integration of a variety of new technologies to create a novel Internet application and social form. In this section, we outline the technologies involved in the Metaverse, as shown in Figure 4.1. The technologies involved in the Metaverse are divided into five aspects, namely network infrastructure, management technology, basic common technology, virtual reality object connection, and virtual reality convergence. The specific explanation is as follows.

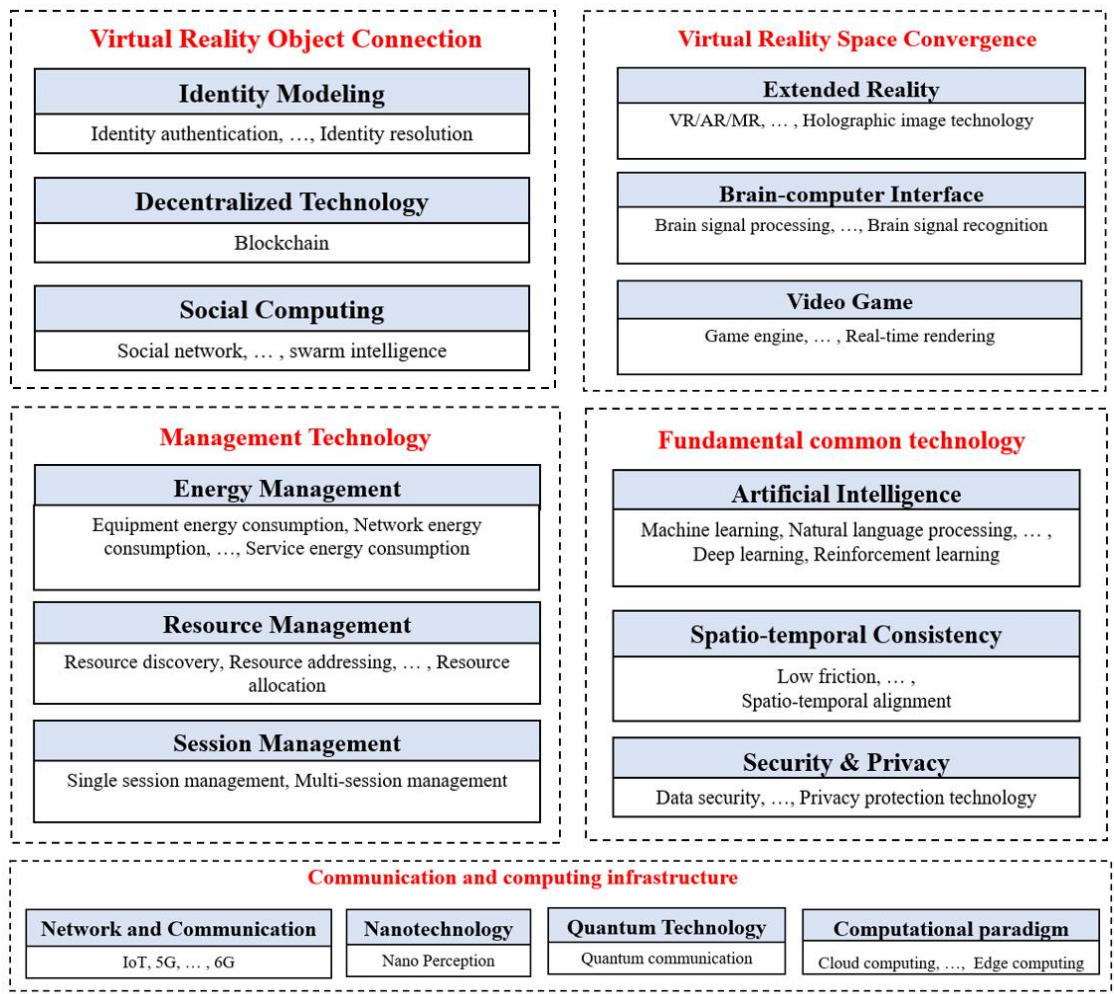

**Figure 4.1 The Technology Roadmap of the Metaverse**

## 4.1 Communication and computing infrastructure

Whether remotely performing large-scale computing tasks, accessing large databases, or providing shared experiences between users, they are inextricably linked to networks and communications [2]. The fifth generation (5G) and the sixth generation (6G) are the communication foundation of the Metaverse. 5G has the advantages of high speed, low delay, ubiquitous network, low power consumption and interconnection of all things, which makes it possible to realize the Metaverse. 6G will break the limitations of time and virtual reality, expand the service objects from humans, machines, and things in the physical world to the "environment" of the virtual world, and realize the cooperation between humans-machines-things-environment by connecting the physical world and the virtual world, providing the network foundation for the Metaverse.

In the 5G and 6G network environment, quantum communication ensures the communication security in the Metaverse. M. Z. Chowdhury et al. [3] proposed that quantum communication provides high security by applying quantum key based on quantum no-cloning theorem and uncertainty principle. Moreover, quantum communication improves the overall security due to the superposition properties of qubits.

In addition, the Internet of Things (IoT) plays a vital role in network infrastructure of the Metaverse. IoT sensing provides users with a completely real, lasting, and smooth interactive experience that bridges the Metaverse and the real world. However, there are still bottlenecks. For example, the imbalance between data explosion and the limited sensing resources. F. Shi et al. [4] proposed the solution of selective perception. Another bottleneck is the poor sensor/actuator performance. To address this issue, H. Ning et al. [4] proposed that Nanotechnology has the ability to improve the performance of sensors/ actuators (e.g., higher sensitivity and selectivity, shorter response time, and longer service life). Therefore, the application of nano materials (e.g., graphene, nanowires, etc.) will provide options for the field of the Metaverse sensing and communication.

Building the Metaverse requires a powerful computing system. Current architectures for computing power cannot yet meet the low-threshold and experience-intensive requirements of the Metaverse. However, cloud computing, edge computing, and other computing paradigms are capable of promoting the development of computing power to some extent [6, 7] and becoming the main infrastructure of the Metaverse.

## 4.2 Management Technology

The fundamental management technologies of the Metaverse provide the environment required for the connection and convergence of the real world and the virtual world, mainly including energy management, resource management, and session management.

The first consideration in energy management technology is the consumption of electric energy by the Metaverse architecture and facilities. Many scholars have proposed methods for energy monitoring. For example, R. Krishnamoorthy et al. [8] proposed a method based on the IoT to monitor load consumption and save energy in an efficient way. G. Bedi et al. [9] developed an Elman recurrent neural network model and an exponential power prediction model to reduce power loss and save costs. The medium- and long-term development of the Metaverse requires the search for more stable and sustainable energy. Energy sustainability is not only the core constraint of the Metaverse, but also an investment opportunity.

An urgent problem to be solved in the context of the Metaverse's resource management technology is how to effectively discover and allocate resources. Researchers are also constantly exploring resource management strategies to provide the foundation for the implementation of the Metaverse. L. N. Nunes et al [10] proposed a resource search and discovery algorithm based on elimination selection to solve the problem of resource search and discovery in heterogeneous environments. R. S. Moorthy et al [11] proposed a new cloud resource discovery mechanism based on a sine-cosine optimization algorithm. Y. Han et al [12] proposed a dynamic resource allocation framework to synchronize the metaverse with IoT services and data.

Session management is the management of the interaction between ubiquitous resources and resource users in heterogeneous networks [5]. In the Metaverse environment, it is important to manage persistent interactions with dynamic characteristics, especially for sessions with multiple resources users. Moreover, the real-time nature of the session is available to increase the user's immersion experience. Park K. Y. et al. [13] discussed how to realize user's high-performance session management in 5G wireless network environment. In addition, Metaverse should also prevent sessions from being attacked. Nadar V. M. et al. [14] studied a defensive approach against destructive authentication and session management attacks. M. Marlinspike et al. [15] developed the sesame algorithm to encrypt asynchronous messages and improve the security factor during the session.

## 4.3 Fundamental Common Technology

The fundamental common technologies of the Metaverse comprise artificial intelligence (AI), spatio-temporal consistency, security and privacy, etc. The details are as follows.

AI algorithms (i.e., machine learning, deep learning, reinforcement learning, etc.) are the "key" to connect the virtual world and the real world. The three elements of AI, namely data, algorithm, and computing power, play a vital role in the establishment and development of the Metaverse. By using artificial intelligence technologies, the Metaverse can safely and freely participate in social and economic activities beyond the boundaries of the real world [16]. Utilizing computer vision, intelligence voice, natural language processing, and other technologies, users have the same visual and auditory sensations as in the real world.

Spatio-temporal consistency is the most fundamental feature of the Metaverse. The ultimate form of the Metaverse is the parallel digital space-time continuum of real human society, so consistent spatio-temporal data is critical for mapping between the real world and the Metaverse. G. Atluri et al. [17] have explored the approaches of spatio-temporal data mining. Moreover, it is also necessary to study spatio-temporal consistency methods, such as time synchronization, target positioning, time registration, and spatial registration [5].

The security and privacy of user data is one of the biggest issues in the real world. With the advent of the Metaverse, the amount and richness of personal data collected are unprecedented. In the future, it is very likely that multiple companies / institutions will work together to build one or more Metaverse, so how to coordinate data with companies / institutions and how to interact data between different Metaverses to ensure the privacy and security of the Metaverse. Z. Zhang et al. [18] surveyed the literature on user access authentication, network situation awareness, dangerous behavior monitoring, and abnormal traffic identification to provide reference for optimizing security and privacy in the Metaverse. B. Falchuk et al. [19] also proposed the privacy issues in the Metaverse.

## 4.4 Virtual Reality Object Connection

The Metaverse will serve as a bridge between the physical and digital worlds, inextricably linked to support for identity modeling, decentralization technology and social computing.

The Metaverse can be simply understood as a network world parallel to the real world. Therefore, as in the real world, individuals entering the metaverse require an identity credential, regardless of whether it is related to the real identity, which identity modeling technology meets their needs. Identity modeling and identity addressing [20] are the bridge between the real world and the Metaverse and will be a very important research field in the age of Metaverse.

Users living in the Metaverse cannot live without social computing. The emergence of the Metaverse will not replace real social relationships with virtual social relationships, but will bring about a new kind of social relationships that are integrated online and offline. Social computing predicts the operation law and future development trend of the Metaverse by studying human behavior and social relationships. In addition, it is easier to collect the location, age, preferences and other information of users in the Metaverse and make a detailed evaluation to better support the society of the Metaverse.

Every part of the Metaverse believes in the concept of decentralization, which needs the help of the underlying technology of decentralization to ensure the security and operation of the Metaverse. Decentralized technology includes blockchain, distributed storage, distributed computing, etc., and the most typical decentralization technology applied in the Metaverse is blockchain technology [21]. B. Ryskeldiev et al. [22] proposed a point-to-point distribution model based on distributed blockchain for the virtual space of mixed reality applications.

## 4.5 Virtual Reality Space Convergence

The Metaverse will profoundly change the organization and functioning of existing society through the integration of virtual reality. To realize the space convergence of the virtual and real world, augmented reality, brain-computer interface, and video game technology are indispensable.

The AR/VR/MR technology is one of the technical pillars in the construction of the Metaverse. Augmented reality (AR) overlays the virtual information at a position based on the detected object through device recognition and assessment (two-dimensional, three-dimensional, GPS, somatosensory, facial and other detected objects) and displays it on the device's screen, allowing the virtual information to interact in real time. Virtual Reality (VR) provides users with a completely immersive experience, making them feel like they are in the real world. It is an advanced and idealized virtual reality system. Mixed Reality (MR) is a new visualization environment that combines real and virtual worlds. In the new visualization environment, physical and digital object coexist and interact in real time. The boundary between VR/AR/MR will blur in the future and become a fusion product. At present, it is the primary interaction technology adopted by the Metaverse to create a highly interactive virtual world for users.

Holographic image is a recording and reproduction technology that presents the real three-dimensional image of an object by optical means. It is the result of the combination of computer technology and electronic imaging technology. It uses coherent light interference to record the amplitude information and phase information of the light wave, and obtain all the information of the object including shape, size, etc. The holographic image is a real three-dimensional image. Users can view the image with naked eyes at different angles without wearing portable devices. With the development of the technology, the boundary between the physical world and the virtual world can be blurred, which will create a solid foundation for the real realization of the metaverse.

Brain-computer interface (BCI) encodes and decodes brain signals in the process of brain activity by accurately identifying brain signals, which can be used by users for operation, such as playing games, typing, etc. BCI connects the human neural world with the external physical world by decoding individual brain signals into commands recognized by computing devices [24], which can realize the space convergence of the virtual world and the real world. R. Abiri et al. [23] have reviewed BCI methods based on EEG. Currently, there is also research on BCI based on AI technologies [24, 25, 26] to accelerate the development of BCI and lay a foundation for space convergence of the Metaverse.

Video game technology is the most intuitive way of presenting the Metaverse. It can not only provide a creative platform for the Metaverse, but also realize the aggregation of interactive content and social scenes. Game engine is the core of video game technology, which refers to the core components of some compiled editable computer game systems or some real-time interactive image applications. The emergence of game engine reduces the level of difficulty for game designers and developers, so that they need not to start with the most basic code. Game engine development drives image development in the metaverse and provides users with an experience that is closer to the real world. Currently developed game engines are listed in Table 4.1.

**Table 4.1 Overview of representative game engines**

| Engine Name | Company | Release Time | Application Game | Link |
|---|---|---|---|---|
| Unreal engine | Epic | 1998 | *War machine*, *Quality effect*, *Ownerless land*, *Absolute survival*, *Escape*, *peace elite*, *Fortress night*, etc. | https://www.unrealengine.com/zh-CN/ |
| Rockstar engine | Rockstar | 1998 | *GTA4, Wild escort* | https://www.rockstargames.com/ |
| Unity engine | Unity Technologies | 2004 | *Glory of the king, Legend of hearthstone, Temple Escape*, etc. | https://baike.baidu.com/item/Unity/10793?fr=aladdin |
| Source engine | Valve | 2004 | *DoTA2, Anti terrorism elite, The fall of Titan, The road to survival* | https://www.moddb.com/engines/source/ |
| IW engine | Infinity Ward | 2005 | *Call of duty, Call of Duty: black action 3* | https://www.moddb.com/engines/iw-engine |
| Frostbite engine | DICE | 2006 | *Battlefield, Medal of honor,* etc. | https://www.ea.com/frostbite |
| Anvil engine | Ubisoft Montreal | 2007 | *Assassin's creed, Prince of Persia 4* | https://www.moddb.com/engines/scimitar |
| Cry engine3 | CRYTEK | 2009 | *Island crisis, Sniper: Ghost Warrior 2, Monster Hunter* | https://www.cryengine.com/ |
| Coscos2D | -- | 2010 | *Defending radish, Fishing expert, My name is Mt* | https://www.cocos.com/ |

# 5. Social Metaverse

The advent of Metaverse will transform the traditional social networks into interactive and immersive 3D virtual social worlds.

## 5.1 Virtual social worlds

The convergence of social networks and virtual reality has enabled the creation of virtual social worlds. These are 3D immersive environments that extend the traditional content-oriented social networks into a fully interactive social simulation. In social virtual world, users are represented by avatars that navigates through the virtual world and socially interact with other users. Users can teleport through different virtual social worlds, participate in events, and even trade real money. Virtual social world are complex social systems that integrate the real social space with the virtual social environment through ubiquitous design of cyber-physical-social-systems (CPSS) [27]. Figure 5.1 shows the enabling technologies that contributed to the development of social metaverse applications. The integration of the physical social space with the virtual social space requires the continuous mapping of social interactions and social events in the virtual social world. The virtual social world must meet the four design requirements. (1) Realism: the requirement for users to sense emotionally immersed in the virtual social world. (2) Ubiquity: the requirement that the virtual social Metaverse be ubiquitously accessible from various devices and locations, and that the user's virtual identities or cyber persona remain connected during transitions within the virtual social world. (3) Interoperability: the ability of the virtual social world to employ standards that allow users to teleport/move seamlessly between different virtual locations in the Metaverse without disconnections and interruptions in their immersive experience. (4) Scalability: The ability of the virtual social world to manage computational power in such a way that a large number of users can interact socially in the Metaverse without experiencing disconnections and interruptions in their immersive experience [28].

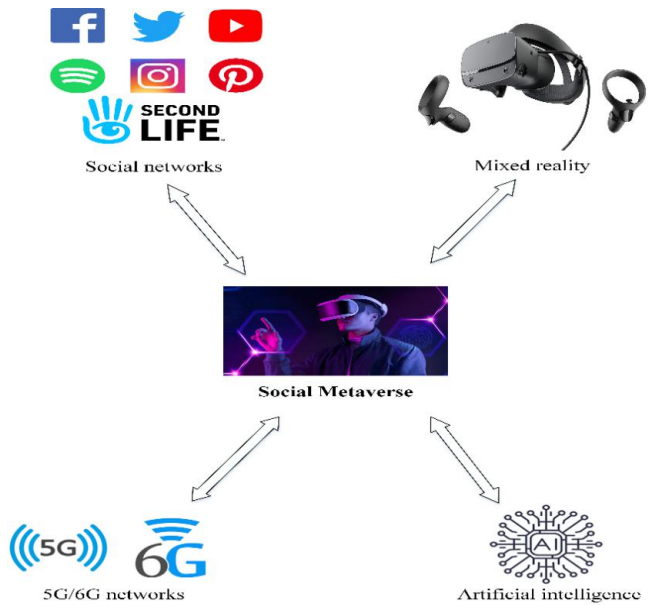

**Figure 5.1 The four enabling technologies for the interactive social Metaverse**

## 5.2 Social privacy in the Metaverse

Digital footprints in the social Metaverse can be tracked to reveal the user's real world identity, and other sensitive information, such as location, shopping preferences and even financial details [29]. The importance of privacy plays a crucial role in shaping the social Metaverse. Applying privacy-preserving schemes is much easier in traditional social networks, as users can decide with whom to share their social media content. On the other hand, such privacy control is not possible in the social Metaverse, as users cannot change the virtual properties of the constructed virtual world, which makes user privacy preservation challenging [30]. Take the following example: If you are navigate a mall in the social Metaverse, and an avatar follows your avatar and records all the things you buy and your travel history, this information can be used to perform social engineering attack that can violate your privacy in the real world. On the other hand, we cannot just turn off who can follow our avatars in the Metaverse as we can do in the traditional social media. Another example of privacy in the Metaverse: You want to have as much privacy in your home that you have created as you do in the real world. However, the current virtual social networks allow other avatars to navigate freely around the map, including your house, and you cannot call the police, as you would do if someone invaded your house.

    One of the proposed solution to address the privacy issues in the social Metaverse is social clone [31]. This involves creating multiple clones of each user in the Metaverse to confuse the attackers that try to stalk individuals in the Metaverse. Such a privacy protection plan may solve some of the problems, but it creates an even bigger problem by allowing users to have multiple representation in the Metaverse, not to mention millions of clones roaming around the Metaverse, that will confuse more than just the attackers. Other privacy protection schemes include disguising users by periodically changing the appearance of the avatar to make it harder for attackers to target specific individuals, and making the avatar temporarily invisible when being tracked [32]. A privacy framework for the social Metaverse can combine multiple privacy schemes where the user chooses to enable a particular privacy option for each situation. Falchuk et al. [31]     proposed a

privacy scheme that combines various privacy techniques such as virtual clone, private copy, mannequin, avatar lockout, avatar disguise, teleport, and invisibility, as shown in Table 5.1. While o other researchers argue that the privacy jurisdiction of the user's country should also be applied in the metaverse, especially with respect to social relationships in the Metaverse. Lo et al. [33] have advocated that Canada's *Personal Information Protection and Electronic Documents Act* (PIPEDA) should be governor privacy law in social virtual networks like Second Life.

**Table 5.1 Privacy-preserving schemes in the social Metaverse**

| Privacy scheme | Description |
| --- | --- |
| Avatar cloning | Creation of multiple avatar clones that look identical to confuse privacy intruders. |
| Disguise | The ability of users to switch multiple disguised avatars. |
| Mannequin | Replacing the avatar with a bot that mimics the user's behavior, and teleport the user to another location in case of suspected stalking. |
| Invisibility | Allow the avatar to become temporary invisible to prevent it from being tracked by privacy intruders or bots. |
| Teleport | The ability of the avatar to instantly teleport to other location in the Metaverse. |
| Private worlds | Allow certain locations of the Metaverse to be occupied as private location only by certain users. |
| Lockout | Some parts of the Metaverse are temporary locked out for private use for some avatars. |

# 6 Metaverse: Hyper Spatiotemporal view

The concept of hyper spatiotemporal is frequently mentioned in ancient Chinese and abroad. In China, the hyper spatiotemporal was first presented in the form of ancient myths and stories. Taoism has elevated hyper spatiotemporal to a theoretical level. As shown in Table 6.1, the classic Taoist Lao-Zhuang literature is a heresy of the traditional view of spatiotemporal [34].

**Table 6.1 Hyper Spatiotemporal Theory in Lao Zhuang Taoism**

| Personage | Theory |
|---|---|
| Laozi | Laozi's "Tao" contains refined spatiotemporal thoughts. |
| Zhuangzi | Zhuangzi's "Tao" breaks the traditional constraints of spatiotemporal with a relativistic way. |
| Liezi | Liezi's "Tao" is related to the detachment of spatiotemporal with the process of "Tao". The shattered spatiotemporal are intuitively reorganized in the form of an imaginary world to reach the "Tao" directly. |

In the West, be it Plato's "Triple World", Aristotle's "Theology", Hegel's "Absolute Idea" and so on, these ancient Western classical theories also reflect hyper spatiotemporal.

**Table 6.2 Western Classical Theory of Hyper Spatiotemporal Theory**

| Personage | Theory |
|---|---|
| Plato | Plato divides the world into three parts: the rational world (the invisible and eternal truth prototype world), the real world (the visible and perishable real material world that imitates the rational world), and the art world (imitates both the rational world and the real world). |
| Aristotle | Aristotle's "Theology" assumes that reason is eternal, without origin and extinction. It is the cause of the movement of all things and has the properties of the hyper spatiotemporal. |
| Hegel | Hegel's "Absolute Idea" assumes that the natural world is the externalization of the "Absolute Idea". The "Absolute Idea" determines the relationship between time and space. It is the form of the spatiotemporal that it develops up to a certain stage. The "Absolute Idea" precedes the spatiotemporal or the hyper spatiotemporal. |
| Sigmund Freud | The dream spatiotemporal mentioned by Freud in the theory of subconscious is a spatiotemporal deviating from reality. |

In the real world, human behavior takes place in physical time and space. Spatiotemporal guarantees the normal course of human behavior, but also restricts human behavior so that it cannot deviate from the real characteristics of spatiotemporal. The Metaverse breaks the constraints of time and space in the real world. The hyper spatiotemporal of the Metaverse can be discussed at two levels of time and space: 1) transcending the constraints of time, returning to the past and reaching the future; 2) transcending of physical space, traversing space, and crossing spatiotemporal in a given period of time.

The hyper spatiotemporal nature of the Metaverse is reflected in in the fact that the metaverse is a spacetime distinct from the real world. It does not stop at the construction of a static digital space, but a virtual space that evolves in parallel with the real dynamic world. It opens up a new habitat for humanity and brings a different experience to users. At present, Metaverse is in a steady development stage, with the blessing of Internet, 5G, VR and other technologies. The Metaverse has given mankind the opportunity to create a holographic digital world in parallel with the traditional real world. In the future, it will be widely used in education, industry and other fields. Detailed applications are listed in Table 6.3.

Table 6.3 The Metaverse hyper spatiotemporal applications

| Field | Application |
| --- | --- |
| Education | Provides students with an immersive educational experience. For example, when learning about planets, display the universe galaxy, which you can zoom in and zoom out so that you can clearly see the texture and features of the universe. When learning about ancient architecture, you can go back to that era and personally experience the construction process and details of the architecture. |
| Industry | Realize virtual verification design, plan and optimize the manufacturing process of the product life cycle, solve the problems of long product trial cycle and unstable manufacturing process, and use highly simulated virtual products for market and actual field testing. |
| Art | More and more exhibits are presented in digital form, allowing museum exhibitions to be extended to more regions. At the same time, a combination of virtual and real exhibits is being used to display precious collections that cannot be touched up close, enabling a variety of interaction methods for people and collections. |
| Medical | It is widely applied in telemedicine, virtual medical, remote care and monitoring, data-driven medical and other fields. Use limited equipment resources to provide highly mobile, digital, real-time, and remote medical services for patient diagnosis, first aid, and nursing care, improve medical efficiency, and realize informatization of the medical industry. |
| Social | Create a parallel social world that is independent of users' real social relationships, increase users' freedom, free them from the constraints of real world, and expand social circles. |

Finally, the conventional order of spatiotemporal exists in the real world. The realm of "Tao" in ancient China and ancient Western surrealism have completely eliminated the attributes of the spatiotemporal, which is ideal and illusory. The Metaverse lies somewhere in between. Despite its hyper spatiotemporal nature, technicians always maintain the spatiotemporal consistency of user interaction. Therefore, there is no need to worry about the confusion of spatiotemporal caused by the Metaverse.

# 7 Forecast of the first application areas

Metaverse may be the first to be applied in the following fields.

| Smart City | As a virtual world parallel to reality, Metaverse uses digital twin technology, which is also an important means to build a smart city. Digital twin technology can digitally map the physical world, fully capture urban data such as people, vehicles, objects, and space, and form a visible, controllable, and manageable digital twin city. It can improve the efficiency of resource utilization, optimize urban management and services, and improve the quality of life of citizens. |
|---|---|
| Entertainment and Game industry | The development of interactive technology has greatly improved the sense of immersion in the gameplay, which can effectively enhance the user experience, playability and enjoyment. |
| Remote office, virtual meeting | Under the influence of the new crown pneumonia epidemic, the importance of telecommuting has been revealed.<br><br>Metaverse can compensate for the limitations of the original remote office model, further improve the functions of remote office and create more opportunities for remote office. |
| Digital sightseeing<br>Digital tourism, digital exhibition | The construction of a metaverse can better realize digital tourism and digital exhibition. The development of digital twin technology and interactive technology allows users to break through the limitations of time, space and other factors, freely visit scenic spots around the world, and get an immersive experience. |
| Psychotherapy | Metaverse can help with psychotherapy.<br>Metaverse can provide the following help:<br>- Construct a virtual and relaxing situation;<br>- Communicate and interact with virtual characters. |
| Education | The construction of Metaverse can help promote children's education, serious games, and preschool education.<br>Metaverse can contribute to education:<br>- Immersive, simulating of realistic scenes to promote the understanding of learning content;<br>- Avoid the harm of reality experiment. |
| Economy | Blockchain technology, decentralization, and the development and rise of new industries within the Metaverse can effectively drive economic development. |

| Culture virtual idols, virtual concerts, and new forms of cultural creation. | As a new social form of society, the Metaverse will give rise to new cultural forms and methods of cultural creation. The development of interactive technology and the further improvement of immersion can effectively promote the development of virtual idols and virtual concerts. |
|---|---|
| Social | Scenario-based social networking, online gatherings, and making friends. Metaverse breaks the boundaries of time and space, and increases the distance between people. People in Metaverse can communicate anytime and anywhere. Metaverse can provide various forms of social interaction. |

Table 7.1 Forecast of the first application areas

# 8 Open Issues

## 8.1 Interaction problem

As a medium between the virtual world and the real world, the interaction technology of the Metaverse needs to meet the following conditions:

・The interactive device is lightweight, convenient to use, wearable and portable.

・The transparency of the interactive medium enables users to ignore the traces of technology and better immerse themselves in the virtual world.

The existing common interactive technologies include: somatosensory technology, XR (VR, AR, MR) technology, and brain-computer interface. XR technology refers to a combination of real and virtual, human-computer interaction environment produced by computer technology and wearable devices. XR as a general term for immersive technologies to merge virtual and real worlds includes virtual reality (VR), augmented reality (AR), mixed reality (MR), and other new immersive technologies that may emerge due to technological advances. Somatosensory technology refers to people interacting directly with their body movements with the devices or environment around them, without using any complex control equipment, and allows people to interact with content immersively.

At present, these two technologies have the problem that the interactive devices are not lightweight and transparent enough, and the cost is high, which makes them difficult to popularize.

Brain-computer interface technology can be divided into three types: invasive, semi-invasive and non-invasive. Invasive refers to the surgical implantation of electrodes into the cerebral cortex. Semi-invasive refers to the implantation of electrodes into the cranial cavity, but outside the cerebral cortex. Non-invasive refers to the interpretation of EEG signals through a wearable device attached to the scalp. Invasive EEG collection is the most accurate, but it has risks such as high surgical risks and rejection of human tissues. The non-invasive method avoids the safety risks of complex surgery, but the signal collection is relatively weak. At the same time, the brain-computer interface also has the problem of being difficult to disseminate.

## 8.2 Computation Issues

Computing power refers to the ability to process data, which is determined by three indicators: calculation, storage, and transmission of data. Computing power is an important productivity in the digital economy era, its facilities are an important support for technological innovation.

Metaverse means a larger number of users, richer network resources and computing resources, and computing power is an important support for Metaverse. The planning of new business formats and Metaverse platform based on cloud computing technology have increased the demand for computing power resources and also provided room for development of computing power. The cloud storage, cloud computing, cloud rendering and other technologies used by Metaverse place high demands on client device performance and server resilience. Metaverse needs to continuously improve processing speed, complexity and power consumption.

## 8.3 Ethical issues

The Metaverse has given people a new identity and create a new, very free space for life and activities. It contains more complicated social relationships. As a next-generation network, Metaverse must control and constrain behavior of users, and establish clear ethical and moral norms to maintain a good and orderly ecological environment of the Metaverse.

The ethical and moral problems of the Metaverse refer to the phenomena that arise in Metaverse due to the absence and confusion of the corresponding moral norms, which conflict with the ethical norms of the real society.

The ethical and moral issues that Metaverse needs to solve:
② Integrity issues-publishing and disseminating false information, fraud；
② Problem of unfavorable atmosphere；
③ Infringement of intellectual property rights.

With the development of Metaverse interactive technology, when the brain's consciousness can be edited, stored, and copied like computer information, the scenes in science fiction movies may no longer be imaginative. At this point, the role of ethics becomes very important. The original code of ethics has been affected, and the formulation of the new code of ethics is lagging behind and cannot keep up with the development of the Metaverse. Therefore, the supervision of the Metaverse should be strengthened, and relevant laws and regulations should be formulated and updated in a timely manner.

## 8.4 Privacy Issues

The Metaverse is closely linked to the real world and corresponds to the real identity. As the construction of a new generation of networks, the Metaverse must take full account of data privacy protection issues, just like the previous network environment.

## 8.5 Cyber-Syndrome

Cyber-Syndrome is a physical, social, and mental disorder caused by excessive use of the Internet [35,36]. With the continuous development of interactive methods, electronic devices have become smaller and more portable. The streamlining of equipment makes people spend more and more time on the Internet. At the same time, the Metaverse is closely connected with the real world. The fusion of virtual and real, and the high degree of immersion of Metaverse make the problem of cyber syndrome even more serious.

## 8.6 Standards and compatibility

As a virtual world that is closely connected to reality and has a multi-dimensional nature, it is necessary to establish standards for the Metaverse. The compatibility and standardization issues of the Metaverse can be divided into two aspects:

Compatibility issues between Metaverses created by different companies.

Compatibility between Metaverses and the real world (including currency compatibility and circulation issues, and handling of legal disputes).

# 9 Conclusion

Metaverse has broad development and application prospects. This paper summarizes the work of different countries and enterprises, collects papers related to Metaverse, introduces the three characteristics of Metaverse's multi-technology, sociality, and hyper spatiotemporality, predicts the first application areas of Metaverse, and discusses its problems and challenge.

Authors:

Huansheng Ning received his B.S. degree from Anhui University in 1996 and his Ph.D. degree from Beihang University in 2001. He is currently a Professor and Vice Dean with the School of Computer and Communication Engineering, University of Science and Technology Beijing, China, and the founder and principal at Cybermatics and Cyberspace International Science and Technology Cooperation Base. He has authored 6 books and over 150 papers in journals and at international conferences/workshops. He has been the Associate Editor of IEEE Systems Journal, the associate editor (2014-2018) and the Steering Committee Member of IEEE Internet of Things Journal (2018-), Chairman (2012) and Executive Chairman (2013) of the program committee at the IEEE international Internet of Things conference, and the Co-Executive Chairman of the 2013 International cyber technology conference and the 2015 Smart World Congress. His awards include the IEEE Computer Society Meritorious Service Award and the IEEE Computer Society Golden Core Member Award. His current research interests include Internet of Things, Cyber Physical Social Systems, electromagnetic sensing and computing. In 2018, he was elected as IET Fellow.

Hang Wang received her B.E. degree from Tianjin Normal University and is currently working on her M.S. degree at the School of Computer and Communication Engineering, University of Science and Technology Beijing, China. Her current research focuses on serious games and the Internet of Things.

Yujia Lin received her B.E. degree from University of Science and Technology Beijing in 2021. She is currently working on her M.S. degree at the School of Computer and Communication Engineering, University of Science and Technology Beijing. Her current research interests include Neuro Linguistic Programming.

Wenxi Wang received her B.E. degree from Ludong University in 2019. She is currently pursuing the M.S. degree at the School of Computer and Communication Engineering, University of Science and Technology Beijing. Her current research interests include social computing and artificial intelligence.

Sahraoui Dhelim received his B.S. in Computer Science from the University of Djelfa, Algeria, in 2012 and his Master degree in Networking and Distributed Systems from the University of Laghouat, Algeria, in 2014. Since 2015 he has been pursuing his PhD at the University of Science and Technology Beijing, Beijing, China. He is an active reviewer in many journals, including IEEE Transactions on Computational Social Systems, IEEE Transactions on Intelligent Transportation Systems and IEEE Transaction on Vehicular Technology. His current research interests include Social Computing, Personality Computing, User Modeling, Interest Mining, Recommendation Systems and Intelligent Transportation Systems.

Fadi Farha (fadi farha@ieee.org) received his M.S. degree in 2017 and currently pursuing a Ph.D. degree at the School of Computer and Communication Engineering, University of Science and Technology Beijing, China. His current research interests include Physical Unclonable Function (PUF), Security Solutions, ZigBee, Computer Architecture, and Hardware Security.

Jianguo Ding (jianguo.ding@bth.se) received his degree of a Doctorate in Engineering (Dr.-Ing.) from the faculty of mathematics and computer science at the University of Hagen, Germany. He is currently an Associate Professor at the Department of Computer Science, Blekinge Institute of Technology, Sweden. His research interests include cybersecurity, critical infrastructure protection, intelligent technologies, blockchain, and distributed systems management and control. He is a Senior Member of IEEE (SM'11) and a Senior Member of ACM (SM'20).

Mahmoud Daneshmand (mdaneshm@stevens.edu) is currently an Industry Professor with the Department of Business Intelligence & Analytics as well as Department of Computer Science at Stevens Institute of Technology, USA. He has published more than 200 journal and conference papers; authored/coauthored three books.